\documentclass{aa}
\usepackage{graphicx}
\begin{document}



\newcommand{\Tef}{T$_{\rm eff}$}


\title{The spectrum of the roAp star HD~101065
(Przybylski's star) in the Li I 6708 \AA\ spectral region
\thanks{Based on observations collected at the European Southern Observatory,
La Silla, Chile (programme 56.E-0640).}}

\author{ A. V. Shavrina \inst{1}, N. S. Polosukhina \inst{2,10},
Ya. V. Pavlenko \inst{1}, \\
  A. V. Yushchenko \inst{3,9}, P. Quinet \inst{4,5},  M. Hack \inst{6}, \\
  P. North \inst{7}, V. F. Gopka \inst{3}, J. Zverko \inst{8}, 
J. \v{Z}hi\v{z}novsk\'{y} \inst{8}, A. Veles~\inst{1}
  }
\offprints{A. V. Shavrina}

\institute{Main Astronomical Observatory of NAS of Ukraine,
         Golosiivo, 03680, Kyiv, Ukraine,
         email:~shavrina@mao.kiev.ua
\and Crimean Astrophysical Observatory,
         Nauchnyj, Crimea, Ukraine,
         email:~polo@astro.crao.crimea.ua
\and Astronomical observatory, Odessa National University,
         65014, park Shevchenko, Odessa, Ukraine
\and Astrophysique et Spectroscopie, Universite de Mons-Hainaut, 
 Place du Parc, 20, B-7000 Mons, Belgium 
\and IPNAS, Universite de Liege, Sart Tilman, B15, B-4000 Liege, Belgium 
\and Department of Astronomy, Trieste Univ., Italy,
          Via Tiepolo 11, 34131 Trieste, Italy
\and Institut d'Astronomie, Universit\'e de Lausanne,
         CH-1290 Chavannes-des-Bois, Switzerland 
\and Astronomical Institute of the Slovak Academy of Sciences,
         05960 Tatranska Lomnica, the Slovak Republic
\and Chonbuk National University, Korea
\and Crimean Branch of INI, Chile 
}
\date{Received November 15, 2002; accepted March 16, 2003}
\authorrunning{A. V. Shavrina et al.}
\titlerunning{Lithium blend 6708 \AA  structure for HD~101065}

\maketitle

\abstract{ We carried out a detailed analysis of spectra of the
unique roAp star HD~101065 
(Przybylski's star) near the resonance doublet Li I 6708~\AA,
using a most complete 
line list including all possible transitions between REE levels of NIST
database.
Our model calculations were performed under two assumptions: blend of
Li and REE lines, and blend of REE lines only. They prove
that Li lines are present in the range $6707.72-6708.02$~\AA\,
and that the resulting Li abundance is 3.1 dex (in the scale
$\log$ N(H) = 12.0), while the isotopic ratio $^{6}Li/^{7}Li$ is near to 0.3.
}


\section{Introduction}
Among the rapidly oscillating Ap (roAp hereafter) stars, HD~101065 is the
most unusual (Przybylski \cite{Prz61}, \cite{Prz66}).
The report of Przybylski in 1976 at the IAU Symposium No~32, showing the very
specific properties of the spectrum of HD~101065, made a very strong
impression on specialists of Ap stars and at once attracted much
attention to this star.
Significant difference between the spectrum of this star and the
spectra of other Ap stars were pointed out by Przybylski : first, the absence
or extreme weakness of lines of elements that appear generally normal or
close to normal in other Ap stars, like iron and some other iron group elements,
then the abnormal strengthening of Rare Earth Elements (REE hereafter) lines.
 Przybylski predicted the presence of a magnetic field for this star. He 
proposed to explain the observed anomalies of elemental abundances
by the stability of the upper atmosphere due a strong  magnetic 
field, which would favour radiative diffusion.
   Wolff and Hagen (\cite{Wolff76}) did find a magnetic field of
about 2.5~KG in this mysterious star. Kurtz and Wegner (\cite{Kurtz79})
began photometric observations of HD~101065 in the seventies and
detected well-defined pulsations with a peak-to-peak amplitude of about
{\bf 0.012~mag} and a period of 12.14 minutes.

It was the first Ap star in which pulsations were found. Now 
this phenomenon is known to be typical of other Ap stars, which form the
group of roAp stars. The first attempts to identify the lines in the spectrum
of HD~101065 were made by Przybylski (\cite{Prz61}, \cite{Prz66}), and 
Warner (\cite{Warner66}). These papers 
mention for the first time the possible presence of the resonance 
lithium doublet.
In the subsequent study of this star, Wegner and Petford
(\cite{Wegner74}) have proposed that it is the coolest of Ap stars. They
considered the very strange spectrum to be the combined result of low
temperature and high surface REE abundances (the abundance excess is about
4~dex).   Beginning with Przybylski, 
investigators suggested that heavy line blanketing changes the atmosphere 
structure very strongly, and that the computation of a realistic model 
atmosphere for this star is complicated by the necessity to take
REE lines blanketing into account. Cowley at al. (\cite{Cow98}, \cite{Cow00})
examined in detail the problem of effective temperature to define the model
atmosphere of HD~101065, and 
have carried out a thorough quantitative analysis (abundances of 54 
elements) of the spectrum in the range 3959-6652~\AA.
Cowley and Mathys (\cite{Cow98}) have noted in their conclusion that the
presence of the third spectrum REE (Pr III, Nd III, Ce III...) is evidence 
for the unusual structure of the atmosphere of this star, which includes a
very thin surface convective zone typical of stars of spectral type F2. One
possible explanation of these extreme abundance anomalies is radiative
diffusion, allowed by the stabilization of the atmosphere by a strong magnetic 
field.
Cowley et al. (\cite{Cow98}, \cite{Cow00}) did not address the question 
of the lithium abundance in the atmosphere of this star.
  The present work is a further investigation of the spectrum of HD~101065
in the region 6675-6735 \AA; the main goal is the analysis of the lithium
blend 6708 \AA\ and estimation of the atmospheric lithium abundance.

The very complicated spectral feature at $\lambda$ 6708 \AA\ in the
observed spectra raises the problem of correct line identification,
especially REE lines, in this region. A comprehensive analysis of REE
lines 
with latest atomic data was performed, and we show that the resonance 
doublet of lithium is
the main contributor to the 6708 \AA\ feature.

\section{Lithium in the spectrum HD~101065}

The presence of lithium in the spectrum of this unique
star was first mentioned by Przybylski (\cite{Prz61}), and this remark
was  noticed by Warner (\cite{Warner66}), who made additional 
observations of this star in Radcliffe Observatory, using the 74" reflector,
in the spectral range 3770 -- 6880 \AA\ (dispersion  6 \AA/mm).
  He found very strong lines of singly ionized REE in the region
5500 -- 6880 \AA, the relative intensities of which are similar to laboratory 
intensities from the tables of Meggers et al. (\cite{Meggers}).  However, 
as mentioned by Przybylski (\cite{Prz66}),the line
Sm II 6707.45~\AA\ can be considered only as part of the lithium
blend, and the main contribution to this blend remains the resonant lithium
doublet Li I 6708 \AA . The first 
estimation of the lithium abundance relative to the solar value was made
by Warner (\cite{Warner66}): [Li] = 2.4 dex. This author also
mentioned the probable presence of $^6{Li}$.

\section{Observations and reduction}

The observations were made by one of us (PN) with the 1.4m CAT at ESO La Silla
in March 1996,
in the framework of the International Project ``Lithium in magnetic CP stars''.
The Coud\'{e} \'{E}chelle Spectrograph was used with resolving
power $R = 100\,000$ and S/N $\geq$ 100.
The usual data reduction (bias and dark subtraction, division by flat-field,
extraction and wavelength calibration) was done by PN during the observing run,
using the old IHAP system of ESO (see North et al. \cite{North98} for more
details).
In Table 1 we give the epoch, exposure time and wavelength coverage of our
observations.
\begin{table*}[hbt]
\small
\begin{center}
\caption{ List of observations for HD~101065.}
\vspace{0.5cm}
\label{t1}
\begin{tabular}{crcccc}
\hline\hline
   &Date& UT& Exp.& HJD & Range   \\
\# &d.m.y&h m& [m] & 2450000+ &    \\
\hline
04 &11.3.96 & 5 11 & 20 & 153.726 & 6675--6735~\AA \\
26 &10.3.96 & 4 41 & 20 & 152.705 & 6120--6180~\AA \\
\hline\hline
\end{tabular}
\end{center}
\end{table*}

\section{Model Atmosphere}
\label{model}
We have computed a model atmosphere 
with 
$T_{\rm eff} =6600$~K, $\log g = 4.2$ and abundances 
of most (54) elements
from Cowley et al. (\cite{Cow00}). 
The other element abundances were adopted from Kurucz (\cite{Kurucz99}). 
A detailed discussion of model atmosphere parameters for HD~101065
can be found in the above mentioned paper, as well as in the paper by
Piskunov and Kupka (\cite{PK01}).

The most recent method for determining fundamental stellar parameters 
through pulsation analysis was applied by Matthews et al. (\cite{Matthews99})
to some roAp stars, including HD~101065. But it needs a photometric
calibration. $H_\beta$ calibration results in $T_{\rm eff} = 7400$~K, 
which is too high, as confirmed by detailed spectral analysis
(see Cowley et al. \cite{Cow00}).
We used the SAM12 program (see Pavlenko \cite{Pav2002},
\cite{Pav2003}), a 
modified version of Kurucz's code ATLAS12 (Kurucz \cite{Kurucz99}).
The essential changes were made in the subroutines accounting for 
opacities.
Namely, the opacity caused by absorption of atomic lines was
taken into account with the Opacity Sampling (OS) technique (Sneden et al. 
\cite{Sneden76}).
 Then, bound-free absorption cross-section of C I, N I and O I atoms 
{\bf in the wide spectral region 30-8880 nm} were calculated by us, 
Pavlenko \& Zhukovskaya \cite{Pav02}, using the data of the TOPBASE 
project (Seaton \cite{Seaton92}). 
{\bf New opacity tables of the bound-free absorption of C I, N I, O I atoms,  
reduced to the format of Hofsaess (\cite{Hof79}) opacity tables,}
are available on the web site of Ya.P.(\cite{Pav03}).
The importance of {\it bf} absorption for formation of spectra of 
different stars are discussed elsewhere (Pavlenko \cite{Pav99, Pav2002}, 
Asplund et al. \cite{Asp2000}).  

The basic advantage of the SAM12 code is the possibility to calculate 
model atmospheres for stars with chemical composition  different from the solar one, as 
well as in ATLAS12. Abundances of elements were set as input data which 
do not change with depth. The microturbulence velocity was set constant
with depth. 

The profile of the line absorption coefficient was defined
by a Voigt function $H(a,v)$; the damping constants were taken from line 
databases or calculated within the frame of the Uns\"old approximation 
(Uns\"old \cite{Unsold55}).
The list of absorption lines of atoms and ions for the range
$\lambda\lambda 40 - 60000$~nm was compiled using the
VALD database (Kupka et al. \cite{vald}); the REE lines data, calculated with
Cowan's code (Quinet et al. \cite{Qui99}), are taken from the DREAM database 
(\cite{dream}). Unfortunately, this database contains lines of
only about one quarter of all rare earth ions. Therefore, in the opacity 
sampling calculation we consider two samples of REE lines. In the first we 
calculated the absorption provided by the VALD and DREAM lines. Then, we
amplified the absorption of DREAM lines by a factor of 10 (in addition to the
VALD lines) to compensate for the incompleteness of the REE line list. 
The
comparison of these two cases shows that REE lines opacity does not affect 
significantly the temperature distribution in our model atmospheres.

Kurucz's models with   $T_{\rm eff} = 6750$~K, $\log g = 4.0$ (Kurucz
\cite{Kurucz99}) and abundances of Cowley et al. (\cite{Cow00}) were also
used in calculations of synthetic spectra.  We 
calculated 6750/4.0/[M] = 6750/4.0/0 model with solar abundances, as 
described previously, and compared them with Kurucz models 
(Kurucz \cite{Kurucz93}). Comparison of temperature structures 
and corresponding  synthetic spectra shows their similarity. 
 As far as our model atmosphere is rather iron deficient,  the 
increase of opacity due to the REE elements may be compensated by the 
weakness of iron lines. 

\begin{figure*}[hbt]
\centering
\includegraphics[width=\textwidth]{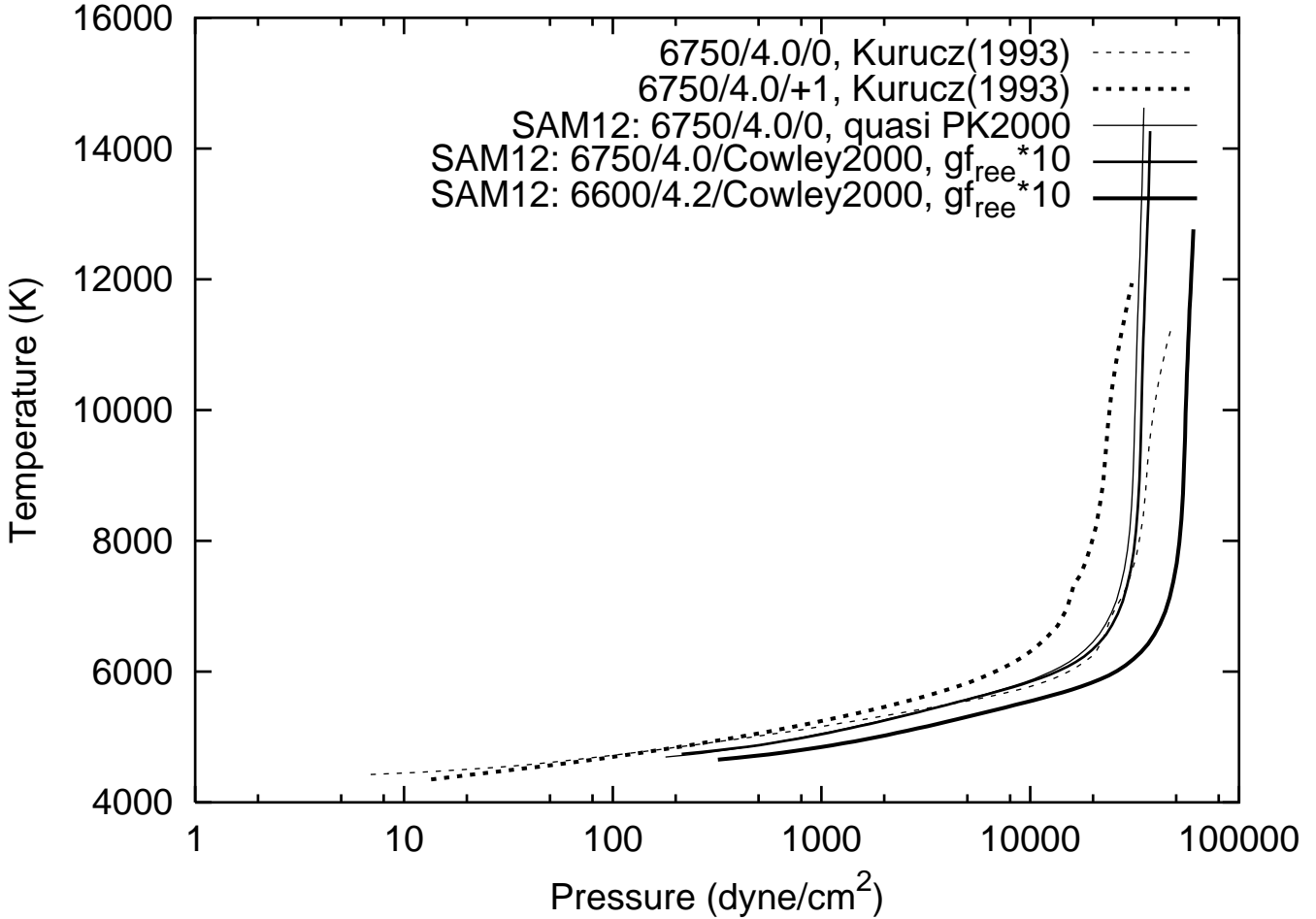}
\caption{T-P relation for model atmospheres considered in this work.
{\bf Two models from the Kurucz's (1993) grid are showed, as well as SAM12
simulation of model atmosphere using Piskunov \& Kupka (2000) approach;
SAM12 model atmosphere computed with Cowley (2000) abundances and REE
bf opacities enhanced by factor 10 of the same \Tef and log g;
structure of model atmosphere \Tef = 6600/4.2 with Cowley abundances and 
enhanced REE opacities is shown also.} See text for more details.
}
\label{fig1}
\end{figure*}

Cowley et al. analysed the spectrum of HD~101065 with ATLAS-9 model with
Opacity Distribution Function (ODF) tables, calculated by the authors on
the basis
of scaled solar abundances to fit better the observed color indices, in 
particular the unusually low c$_{1}$ index. The model atmosphere for this star 
was also computed by Piskunov and Kupka (\cite{PK01}), also through ODF method, 
with a 2.25 dex Fe overabundance and a 2.49 dex Ca excess to simulate unknown
opacity
in ODF tables. These authors used the abundances of Cowley et al.(\cite{Cow00})
for model 
computation with these ODF. We have followed this approach for OS method, by
computing line opacity due Fe I and Ca I lines with factors for gf- values of 
177.8 (2.25 dex) and 308 (2.49 dex) respectively. In Fig.~1 we compare 
the T-P relation for model atmospheres with $T_{\rm eff}={\bf 6750}$ and 
$\log g =4.0$ of Kurucz (solar abundance and $[M] = +1$) and ours with 
Cowley's abundances, calculated in two ways -- one OS model, similar to the 
ODF model of Piskunov \& Kupka {\bf (SAM12:6750/4.0/0, analogue PK2000)}, 
and our two OS model atmospheres for 
$T_{\rm eff}={\bf 6750}$ and 6600~K and $\log g=4.0$, with gf(REE) lines 
enhanced by a factor of 10{\bf(SAM12:6750/4.0/0/Cowley 2000, $gf_{ree}$*10
and SAM12:6600/4.2/0/Cowley 2000, $gf_{ree}2$*10)}.
In general, the photospheres of models with higher opacities are 
shifted upwards, into the layers of lower pressure; however, the differences 
in the T-P relation are rather small.

It is worth noting, that our OS model atmospheres 6750/4.0/0 and
6750/4.0/+1 agree with corresponding Kurucz's ODF models within 50
K in the line forming regions. Therefore, our OS model atmosphere with
``metal rich''
opacities should follow the Piskunov \& Kupka (\cite{PK01}) ODF model.

To analyze the Li blend, we used an atmosphere model with $T_{\rm eff}=6600$,
$\log g=4.2$. The resulting abundances were compared for this model and for
the OS model, which is similar to that of
Piskunov \& Kupka for the same parameters, both for Cowley's 
abundances, and for Kurucz's model for solar abundances. They are cited and 
discussed in the next section.

\section {Li blend analysis and REE lines contribution to the 
 6705.75-6708.75~\AA\ spectral region}

 We calculated the synthetic spectra in the range
 6705.75-6708.75 \AA\ 
 with the help of the STARSP ( Tsymbal 
\cite{Tsymb96}) and ROTATE codes (see Shavrina et al. \cite{Shavr00}), 
both developed by V. Tsymbal, and of the URAN code (Yushchenko \cite{Yu97}). 
All cited codes are based on Kurucz' modules for 
synthetic spectra ( "SYNTHE", \cite{Kurucz93}). 
The URAN program allows to select the elemental abundances automatically,
and applying it to the available spectral ranges (6120-6180, 6675-6735 A),
one obtains results which are close to those of Cowley et al.(\cite{Cow00}).

\begin{table*}
\small
\begin{center}
\caption{The list of lines used for spectra calculations in the range
 6705.75 - 6708.75 \AA}
\label{t2}
\begin{tabular}{|c|c|c|c|c|c|c|c|c|c|c|}
\hline
 el&$\lambda$,\AA& E"&$\log gf$&source&&el&$\lambda$,\AA&E"&$\log gf$&source \\ 
\hline 
*Dy II& 6705.727& 2.078&-2.68&      &&* Gd II& 6707.462& 3.270& -1.98&     \\
*Nd II& 6705.891& 3.269&-2.70&      &&  Sm II& 6707.473& 0.930& -1.48&VALD \\
  Yb II& 6705.965& 5.856&-3.04& DREAM&&  Yb II& 6707.603& 6.651& -1.38&DREAM\\
  Ce II& 6706.051& 1.840&-0.95& DREAM&&* Sm II& 6707.648& 1.746& -1.27&     \\
Tm II&6706.150&5.322&-1.07&DREAM&& Nd II&6707.755&0.170&-3.55&DREAM[Quinet] \\
Tm II&6706.262&3.955&-2.36&DREAM&&$^{7}$Li I&6707.756&0.000&-0.427&[Smith et al.]  \\
Ce II&6706.307&3.195&-2.40&DREAM&&$^{7}$Li I&6707.768&0.000&-0.206&  -"-      \\
*Pr III&6706.492&3.104&-1.28&   &&    *Sm II&6707.779&2.037&-2.68&           \\
 Pr III&6706.705&0.550&-1.64&DREAM&&$^{7}$Li I&6707.907&0.000&-0.932&[Smith et al.]  \\
*Nd II&6706.738&2.868&-2.48&      &&$^{7}$Li I&6707.908&0.000&-1.161&  -"-   \\
*Sm II&6706.789&1.586&-2.00&      &&$^{7}$Li I&6707.919&0.000&-0.712&  -"-   \\
*Sm II&6706.807&1.874&-1.78&      &&$^{6}$Li I&6707.920&0.000&-0.478&  -"-   \\
 Tm II&6706.906&4.908&-2.47& DREAM&&$^{7}$Li I&6707.920&0.000&-0.931&  -"-   \\
*Nd II&6706.922&3.211&-0.88&      &&$^{6}$Li I&6707.923&0.000&-0.179&  -"-   \\
*Nd II&6707.015&1.490&-1.88&      &&Nd II&6708.030&1.522&-1.13&DREAM[Quinet] \\
*Nd II& 6707.033& 2.222&-3.68&    &&$^{6}$Li I&6708.073&0.000&-0.304&[Smith et al.]        \\
 Ce II& 6707.121& 1.255&-3.76& DREAM&&Ce II&6708.077&2.250&-2.57 & DREAM    \\
*Dy II& 6707.153& 3.292&-1.27&      &&*Er II&6708.088& 3.155& -2.58 &      \\
*Dy II& 6707.266& 2.890&-1.28&      && Ce II&6708.099& 0.701& -2.12 & DREAM \\
*Sm II& 6707.342& 0.884&-2.00&      &&*Nd II& 6708.400&  3.192& -2.48 &      \\ 
*Er II& 6707.418& 3.482&-1.44&      &&*Nd II& 6708.458&  3.536& -1.08 &      \\ 
*Nd II& 6707.433& 1.499&-2.17&      &&*Nd II& 6708.629&  0.746& -4.58 &      \\ 
*Nd II& 6707.453& 2.880&-3.18&  \\
\hline
\end{tabular}
\\ $^{*}gf$-values  were estimated by us from best fitting with
    observed spectrum.
\end{center}
\end{table*}

The line list for spectra calculations was compiled from DREAM database
(\cite{dream}) and VALD (Kupka et al. \cite{vald}).
When there was a choice, the preference was given to $gf$ values from DREAM. 
The wavelength and gf-values for lithium lines ($^7$Li and $^6$Li) were taken
from the paper of Smith et al. (\cite{Smith98}). 
In addition, we have calculated wavelengths of REE lines, singly and 
doubly ionised, using the data of REE energy levels from the NIST database 
(\cite{nist}) 
according to selection rules ( $ \delta J $ = 0, $ \pm $ 1) and combining
even and odd levels of energy (odd-even or even-odd).
All possible transition between REE energy levels of NIST were included in 
the line list for synthetic spectra calculations.
For 
additional to VALD and DREAM lines which are
rather distant from Li lines and do not contribute noticeably to the 6708
\AA\ blend ( see Table~4), $gf$-values were matched for better agreement of
observed and calculated spectra.

$gf$-values for the two lines nearest to the lithium lines (Tables~2, 4),
Nd II 6707.755 \AA\  and
Nd II 6708.030 \AA, were estimated by P. Quinet with Cowan's code
using for the second line the trial identification of the top level, which was
unspecified in the NIST list of REE energy levels.
This line, Nd II 6708.030 \AA\, forms a blend with the strong line
Ce II 6708.099 \AA, whose $gf$ value is contained in the DREAM
database.
In the region studied, there is a rather pure line Ce II 6706.051 
\AA\ .
DREAM value of $\log gf$ = -0.95 for this line provides Ce abundance
$\log$ N(Ce)= -7.63, which almost coincides with the estimate of Cowley et al.
(\cite{Cow00}), $\log$ N(Ce) = -7.60. This Ce abundance value results in
estimate $\log$ N(Nd) = -7.97 from the line Nd II 6708.030 \AA\ with
$\log gf = -1.13$. 
This estimate differs from the value of Cowley et al.
(\cite{Cow00}) $\log$ N = -7.65 ($\pm$0.28) by 0.32~dex, 
but it only slightly exceeds the error value.
Therefore, Quinet's estimate for the gf value of Nd II 6708.030
seems a reasonable
one. However, its blending with the
$^6$Li lines leads to some uncertainty in the
estimation of the lithium isotopic ratio.
The line  of Sm II, $\lambda$ 6707.779, almost coincides with the centre
of the lithium blend, but unfortunately, the calculations for Sm II
are not possible because the energy matrix dimensions exceed the permitted
sizes for Cowan's code.
We tried to calculate the blend 6708 \AA\  with  Sm II line absorption
$\lambda$ 6707.779, instead of Li lines, adjusting the $gf$ value
for better agreement with the observed blend. In this way we could estimate
an upper limit to $\log gf$ for this line, $-0.63$.
The resulting fit was bad (see Fig. 2). Only the inclusion of $^{7}$Li and
$^{6}$Li lines (see Table 2) with the ratio $^{6}Li/^{7}Li = 0.3$ allowed us
to represent the observed profile of blend rather well (see Fig.2). Our
estimate of $\log gf$ for Sm II line $\lambda$ 6707.779 in this case
is equal to -2.68. 
We should repeat, that our estimate of lithium isotopic
ratio is influenced by blending with the Nd II line 6708.030 \AA.

Table 2 contains the line list used in model calculations. The wavelengths
of lines marked by asterisk were calculated on the basis of NIST energy levels
(\cite{nist}).
Corresponding $gf$-values were estimated by us using element abundances
from Cowley et al. (\cite{Cow00}), corrected for some REE elements 
(see Table 3) on the basis of lines with  known $gf$-values (from DREAM and
VALD lists). 
We used a value of microturbulent velocity  V$_t$ = 2 km/s to compensate 
for magnetic line broadening. This value was selected by us for better
agreement of computed and observed line profiles. 

In Table~3 we present our results for Kurucz's and our models
with gf(REE)*10 and OS model (PK in the table) which is similar to that of
Piskunov and Kupka (\cite{PK01}), and compare them with data of
Cowley et al.( \cite{Cow00}) for V$_t$ = 2~km/s. 
A better agreement was achieved for Ce II on the basis of several
Ce II lines with DREAM gf-values, the lithium abundances being
almost the same for all models used in this work. We were forced to reduce
the Nd II and Sm II abundances taken from Cowley et al. ( \cite{Cow00})
when  we fitted the observed line profiles with theoretical ones.
These values (see Table~3), based only on one line of each element,
6708.030 \AA\ of Nd II and 6707.473 \AA\ of Sm II,
can not be considered as reliable abundance estimates.

We should note that the distinctive feature of this star among other roAp 
stars is the predominance of numerous strong Ce II lines in the
spectrum, which 
betrays 
the extreme overabundance of this ion. 
Not accounting for DREAM line list of Ce II can result in
overabundances for some REE and other elements.

\begin{table*}
\begin{center}
\caption{ Element abundances for modelling  6705.75-6708.75 \AA\ spectrum }
\vspace{0.5cm}
\label{t3}
\small
\begin{tabular}{|@{}c@{}|c|c|c|c|c|c|}
\hline
 el &$^7$Li only&\multicolumn{3}{|c|}{$^6Li/^7Li$ = 0.33}& \\ \hline
 &6600/4.2 &6600/4.2&6750/4.0&6600/4.2(PK)& 6600/4.2(Cowley et al.(\cite{Cow00}) \\
 \hline
 Li I & -8.90& -8.90&  -8.80& -8.83&     -       \\ \hline
 Ce II &          -7.63& -7.63&  -7.65& -7.63&     -7.60+-0.26 \\ \hline
 Pr III&          -7.55& -7.55&  -7.60& -7.65&     -7.46+-0.16 \\ \hline
 Nd II&           -7.97& -8.07&  -7.87& -8.07&     -7.65+-0.28 \\ \hline
 Sm II &-8.32& -8.32&  -8.25& -8.42&     -7.75+-0.19 \\ \hline
\end{tabular}
\end{center}
\end{table*}

\begin{figure*}
\centering
\includegraphics[width=\textwidth]{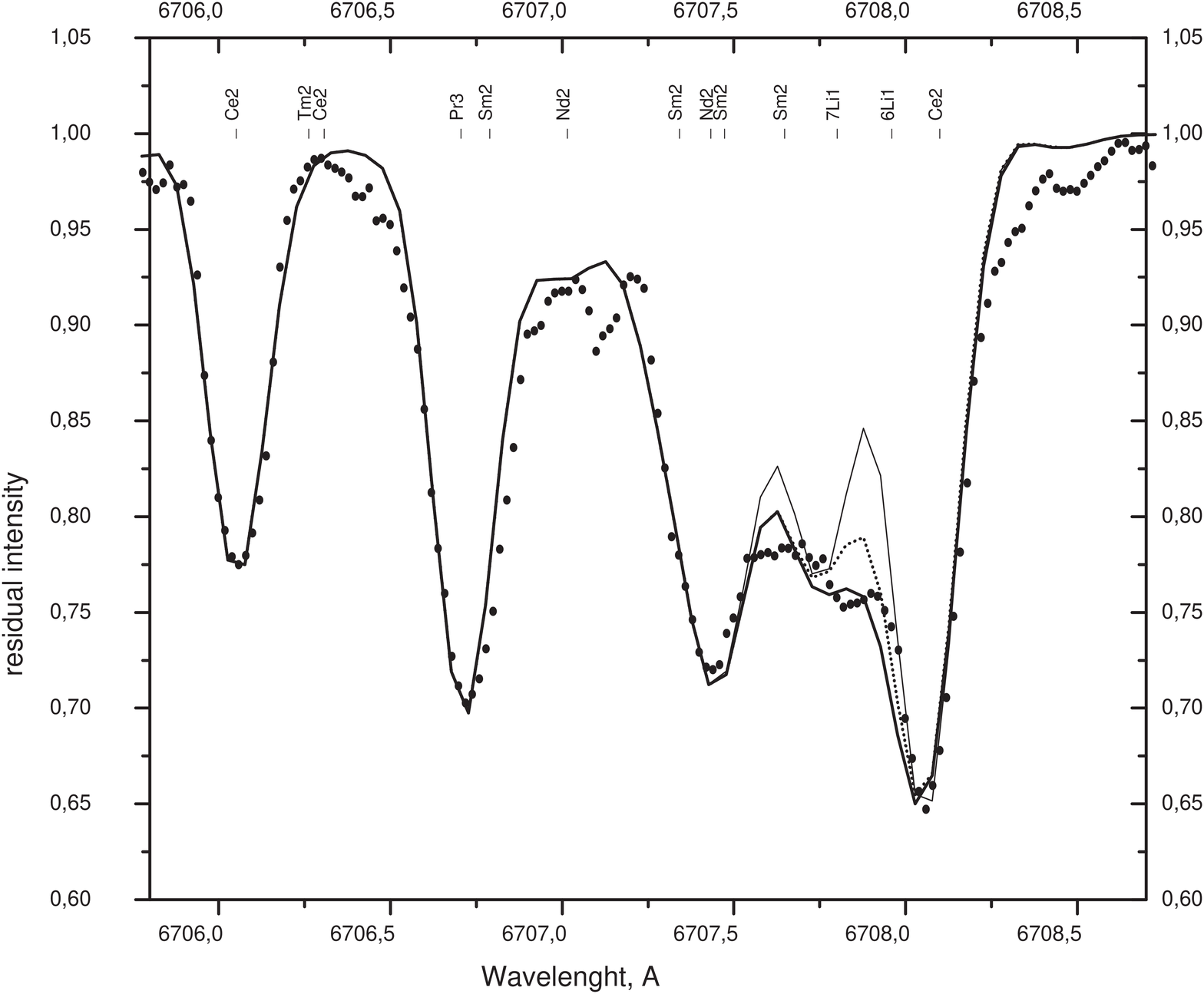}
\caption{The fitting of observed and calculated spectra of
HD~101065 near 6708~\AA~~:
dots: observed spectrum; dashed line: calculated spectrum taking into
account lines of the main isotope $^{7}$Li only; thick line: spectrum with
the ratio $^{6}Li/^{7}Li = 0.3$.
The thin line corresponds to a spectrum computed without Li but with the
Sm II line 6707.799~\AA.
The positions of those lines which are the main contributors in absorption,
are marked at the top of the figure.
}
\label{fig2}
\end{figure*}

\begin{table*}
\begin{center}
\caption[]{ The shares of main absorption contributors in the range 6707.60 -
  6708.16 \AA}
\vspace{0.in}
\label{t4}
\small
\begin{tabular}{|@{}c@{}|c@{}|@{}c@{}|@{}c@{}|@{}c@{}|@{}c@{}|@{}c@{}|@{}
c@{}|@{}c@{}|@{}c@{}|@{}c@{}|@{}c@{}|@{}c@{}|@{}c@{}|@{}c@{}|@{}c@{}|@{}c
@{}|@{}c@{}|@{}c@{}|@{}c@{}|@{}c@{}|@{}c@{}|@{}c@{}|@{}c@{}|@{}c@{}|@{}c@{}|
@{}c@{}|@{}c@{}|@{}c@{}|@{}c@{}|@{}c@{}|@{}c@{}|@{}c@{}|}
\hline
$El$   & $\lambda$,\AA      
&.60&.62&.64&.66&.68&.70&.72&.74&.76&.78&.80&.82&.84&.86&.88&.90&.92&.94&.96&.98&.00&.02&
.04&.06&.08&.10&.12&.14&.16\\ \hline \hline
  Sm\,{\sc ii}  & 6707.648 &6&21&20&20&12&6&2&1&&&&&&&&&&&&&&&&&&&&& \\ \hline
Nd\,{\sc ii}  & 6707.755 &&&&&&1&1&2&2&1&1&&&&&&&&&&&&&&&&&& \\ \hline
$^{7}$Li\,{\sc i}   & 6707.756 &1&2&3&3&4&5&6&7&7&6&6&5&4&2&1&1&1&&&&&&&&&&&& \\ \hline
$^{7}$Li\,{\sc i}   & 6707.768 &1&3&4&4&5&7&9&11&11&11&10&9&7&4&3&2&2&1&&&&&&&&&&& \\ 
\hline
$^{7}$Li\,{\sc i}   & 6707.907 &&&&&&&&&&&1&1&1&1&2&2&2&2&2&1&1&1&&&&&&& \\ \hline
$^{7}$Li\,{\sc i}   & 6707.908 &&&&&&&&&&&1&1&1&1&1&1&1&1&1&1&1&&&&&&&& \\ \hline
$^{6}$Li\,{\sc i}   & 6707.919 &&&&&&&&&&1&1&1&2&2&3&3&3&3&3&3&2&1&1&&&&&& \\ 
\hline
$^{7}$Li\,{\sc i}   & 6707.920 &&&&&&&&&&&1&1&1&1&1&2&2&2&2&2&1&1&&&&&&& \\ 
\hline
$^{6}$Li\,{\sc i}   & 6707.920 &&&&&&&&&&&1&1&1&1&2&2&2&2&2&2&1&1&&&&&&& \\ 
\hline
$^{6}$Li\,{\sc i} & 6707.923 &&&&&&&&&&&1&2&2&2&3&3&3&4&3&3&3&1&1&1&&&&& \\ \hline
Nd\,{\sc ii}  & 6708.029  &&&&&&&&&&&&&&&&&&2&8&26&62&75&71&55&31&12&3&&\\ \hline
$^{6}$Li\,{\sc i}&6708.073&&&&&&&&&&&&&&&&&&1&1&2&2&2&3&3&4&4&3&2&1 \\ \hline
Ce\,{\sc ii}  & 6708.077  &&&&&&&&&&&&&&&&&&&&&&&&1&1&1&&&\\ \hline
Ce\,{\sc ii}  & 6708.099  &&&&&&&&&&&&&&&&&&&&&1&3&8&21&37&46&28&13&5\\ \hline

\hline
\end{tabular}
\label{tb3}
\end{center}
\end{table*}

In Table 4 we show the shares of the main contributors in total absorption
(in percent) in each wavelength with a 0.02 \AA\ step, as they are 
calculated by the STARSP code, for the lithium doublet range
$6707.60 - 6708.16$~\AA\,
without instrumental smoothing, for the model atmosphere 6600/4.2 with 
$V_{t}=2\,{\rm km\, s^{-1}}$ and $^{6}Li/^{7}Li$ = 0.3.

Reyniers et al. (\cite{reyn02}) draw attention to a possible
error in Li line identification (instead of Ce II line 6708.03)
in the 
spectra of s-process enriched post-AGB stars. In our high resolution spectra,
these two lines are sufficiently far apart to be distinguished,
as can be seen in Figure~2 and Table~4.

\section {Discussion of the results}
Since the discovery of the star HD~101065 by Przybylski, the debates
relating to the effective temperature of this star have not settled yet. 
Apparently, the reason lies in the large blanketing effect due to strong 
absorption in REE lines, identified and unidentified. Incompleteness of 
data for these lines does not allow to take into account correctly this 
effect in model atmosphere calculations.
   Cowley et al. (\cite{Cow00}) have clearly shown a large excess of
REE abundances (~4 dex ) and deficiency of iron peak elements ( Fe, Ni ), while
Co has appeared in excess of 1.5 dex. The list of REE lines in this work 
included new identified lines in the spectral range 
bf 3959-6652 \AA.
Our results include the range 6670 -- 6735 \AA \, where
lies the remarkable feature with the resonant lithium doublet line 
6708~\AA.

We have carried out detailed calculations of the blend 6708 \AA
(6705.75 - 6708.75 \AA), using the REE list of atomic data from the
DREAM and VALD databases.
The evidence for presence of lithium in  the spectrum of HD~101065 is most
probable, as shown by the excellent fit of synthetic spectra which include
the Li lines, while the synthetic spectra without lithium (only REE lines, 
Sm II 6707.799) clearly fail to achieve a good match.
All possible transitions between REE energy levels of NIST were included in 
the line list for synthetic spectra calculations.
The presence of other REE lines near 6708 \AA, that would not be included
in our list remains improbable.
All transitions near 6708 \AA\ are due low energy levels, and
low energy REE levels that would not belong to NIST are unlikely. The only
line which might mimick the Li lines is Sm II 6707.799 \AA\ .
But we showed that absorption 
in this line is not large enough to account for the observed spectral feature,
while a perfect fit is obtained when accounting for the Li line components,
which are distributed on a rather wide range 
(see Tables~2 and 4).

In our work the choice of parameters of synthetic spectrum near the
line Li I 6708 \AA\ is described in detail taking into account the
REE lines with known and unknown $gf$ values. The variants of the calculations,
with and without lithium lines, are considered, which prove the presence
of a lithium doublet 6708 \AA\ in the spectrum of HD~101065 
with a lithium abundance of 3.1 dex ( in the scale $\log$ N(H) = 12.0~). 
The observed blend is best represented at the isotopic ratio $ ^6Li / ^7Li $
= 0.3. 
The latter value can be considered as an upper limit in the current
state of atomic data; 
more accurate estimation of the lithium isotopic ratio would require better 
spectral resolution and better $gf$-values for REE lines.

We carried out the calculations with Zeeman and Paschen-Back
splitting of the Li line. For H=2300~G, the Paschen-Back effect is negligible 
(this is common for magnetic fields lower than 4000~G -- see Shavrina et al.
\cite{Shavr00}) and 
can not induce any additional asymmetry. Zeeman broadening is also unimportant,
due to the significant separation (0.15 \AA) of the components of the Li
doublet.
On the basis of our spectra calculations in two wider spectral ranges, 
$6120 -6180$~\AA\ and $6675 - 6735$~\AA, the value V$_t = 2$~km/s was adopted,
which must compensate for the magnetic broadening of other lines. 
We have also considered the possible binarity of HD~101065.
This problem was discussed by Hubrig
et al. (\cite{Hubr00}). They noticed that none of the 14 roAp considered,
including HD~101065, is known to be a spectroscopic binary (although several
are members of wide visual binary systems); therefore the 6708~\AA\ feature
cannot be due to a secondary component of this star .

Warner's (\cite{Warner66}) estimate of the lithium 
abundance for this star 
from a spectrum  with 0.3 \AA\ resolution was 2.4~dex relative to the solar 
value, that  is, 3.3~dex  relative to hydrogen. He also assumed the presence of 
$^6$Li on the surface of the star.
The good agreement between our results and those of Warner (\cite{Warner66})
is encouraging.
Warner  had tried to explain the large excess of 
rare earth elements and the peculiar structure of the atmosphere by
dragging out r-process elements from deeper to surface layers of the star.
However, the high lithium content observed, 3.1 dex ( on the scale where
$\log N(H)=12$~), 
is not {\bf unexpected} for normal F stars, which are separated on high and low Li 
abundances  (Chen et al., \cite{Chen01}),
while Warner's assumption would imply complete mixing and, therefore,
complete lithium destruction. Thus the radiative diffusion mechanism seems a
much more promising explanation, especially as the magnetic field, reaching
2300~G (Cowley et al., \cite{Cow00}), would make it easier. On the other hand,
the isotopic  ratio $^6Li/^7$ Li = 0.3  might be considered as indicating
spallation reactions on the stellar surface, though a combination of radiative
and ambipolar diffusion might perhaps lead to the same result.

   The Hipparcos measurements have confirmed that the star is a dwarf 
or a subgiant: using its parallax ($7.95\pm 1.07$~mas) and the adopted
$T_{\rm eff}=6750$~K, we obtain $\log g = 4.00\pm 0.13$ by interpolation in
evolutionary tracks, assuming a normal evolution and a solar global composition.
However, its spectrum is reminiscent of S-stars (except for Li !), whose
atmospheres are enriched by fresh synthesized
material from deep layers. Whether the extreme anomaly of the spectrum of this
interesting star is just another case of radiative diffusion, or requires
a completely new explanation, remains to be seen.

\begin{acknowledgements}

The data from Kurucz's CDROM 23, NASA ADC, VALD2, NIST and DREAM, were used
and  we thank the administrations of these databases, provided access to
data through INTERNET.\\
N.S. Polosukhina thanks the Consortium of Physics of the Trieste University
for financial support.  A.Yushchenko was supported by the grant of Post-Doc 
Program, Chonbuk National University (2002).
We thank our referee, Dr. M. Gerbaldi, for valuable critical comments
that have helped us {/bf to improve} the paper.

\end{acknowledgements}


\begin{thebibliography}{}

\bibitem[2000]{Asp2000}  Asplund M., Gustafsson B., Lambert D.L.,
Rao N.K., 2000, A\&A, 353, 287


\bibitem[\textit{http://www.umh.ac.be/$\tilde{\ }$astro/dream.shtml}]{dream}
     Bi'emont E., Palmeri P. and Quinet P.,
    D.R.E.A.M. Database on Rare Earth at Mons.Univ.,
    \textit{http://www.umh.ac.be/$\tilde{\ }$astro/dream.shtml}

\bibitem[2001]{Chen01} Chen Y.Q., Nissen P.E., Benoni T. and Zhao G., 2001,  
                       A\&A 371, 943

\bibitem[1998]{Cow98} Cowley C.R. and Mathys G., 1998,  A\&A 339, 165

\bibitem[2000]{Cow00} Cowley C.R., Ryabchikova T., Kupka F., Bord J.D.,
          Mathys G., Bidelman W.P.: 2000,  MNRAS 317, 299

\bibitem[1979]{Hof79} Hofsaess D. 1979, Atom.data.Nucl.data Tab.,24, 285

\bibitem[2000]{Hubr00} Hubrig S., Kharchenko N., Mathys G.and
                    North P., 2000, A\&A 355, 1031

\bibitem[1999]{vald} Kupka F., Piskunov N.E., Ryabchikova T.A.,
         Stempels H.C., Weiss W.W., 1999, A\&AS 138, 119

\bibitem[1979]{Kurtz79} Kurtz D., Wegner G., 1979,  ApJ 232, 510

\bibitem[1993]{Kurucz93} Kurucz R. L., 1993, CD ROM  N 1-23, Cambridge, 
     MA: Smithsonian Astrophys. Obs. 

\bibitem[1970]{Kurucz70} Kurucz R. L., 1970,  Smithsonian
         Astrophys. Obs. Spec. Rept. N 309, 1

\bibitem[1999]{Kurucz99} Kurucz R. L., 1999,
       \textit{http://cfaku5.harvard.edu/}

\bibitem[1999]{Matthews99} Matthews J.M., Kurtz D.V. and 
                      Martinez P., 1999, ApJ 511, 422
\bibitem[1961]{Meggers} Meggers W. F., Corliss C.H., Seribnez B.F., 1961,
            Nat. Bur. Stand., Mon.32

\bibitem[1998]{North98} North P., Polosukhina N., Malanushenko V., Hack M., 
                        1998,  A\&A 333, 644

\bibitem[2001]{nist}
    Martin W.C., Sugar J., Musgrove A., 2001,  Energy Levels
Database,  {\em http://physics.nist.gov/cgi-bin/AtData}

\bibitem[1999]{Pav99} Pavlenko Ya. V., 1999, Astron. Rept. 43, 94
{\bf
\bibitem[2003]{Pav02} Pavlenko Ya.V, Zhukovskaya S., 2003,
          Kinematics and Physics of Celestial bodies. 19, 29.

\bibitem[2003a]{Pav2002} Pavlenko Ya.V., 2003a, Proc. of IAU210, Uppsala, 
    ed. N. Piskunov, in press (astro-ph  0209022)

\bibitem[2003b]{Pav2003} Pavlenko Ya. V., 2003b, Astron. Rept., 47, 59

\bibitem[2003c]{Pav03} Pavlenko Ya. V., 2003c,
        (http://www.mao.kiev.ua/staff/yp/Results).
}

\bibitem[2001]{PK01} Piskunov N. and Kupka F., 2001, A\&A 547, 1040

\bibitem[1961]{Prz61} Przybylski A., 1961, Nat 189, 739

\bibitem[1966]{Prz66} Przybylski A., 1966, Nat 210, 20

\bibitem[1999]{Qui99} Quinet P., Palmeri P., Bi'emont E., 1999,
                   JQSRT 62, p. 625

\bibitem[2002]{reyn02} Reyniers M., Van Winckel H., Bi'emont E., Quinet P.,
               2002, A\&A, 395L

\bibitem[1992]{Seaton92} Seaton M.J., 1992, Rev. Mex. Astron. Ap. 23, 180

\bibitem[2000]{Shavr00} Shavrina A.V., Polosukhina N.S., Tsymbal T.,
        Khalack V.R., 2000, Astron. Rept. 44, 235

\bibitem[2001]{Shavr01} Shavrina A.V., Polosukhina N.S., Zverko J.,
        Mashonkina L.I., Khalack V., Ziznovski J., Hack M., Tsymbal V.,
        North P., Vygonec V.V., 2001, A\&A 372, 571

\bibitem[1998]{Smith98} Smith V.V., Lambert D.L., Nissen P.E., 1998,
                       ApJ 506, 405

\bibitem[1976]{Sneden76} C. Sneden, H. Johnson, B. Krupp, 1976,
                 ApJ 204, 281

\bibitem[1996]{Tsymb96} Tsymbal V., 1996,  M.A.S.S.,
            ASP Conf.Ser., vol. 108, ed. S. Adelman et al., p.\ 198 

\bibitem[1955]{Unsold55} Unsold A., 1955,  Physics der Sternatmospharen,
        2nd ed., (Springer: Berlin)      

\bibitem[1997]{Yu97}  Yushchenko A.V., 1997,
        Proceedings of the 20th Stellar Conference of the
       Czech and Slovak   Astronomical Institutes,
          Brno, Czech Republic.
         ed.\ J. Dusek. ISBN 80-85882-08-6. Brno 1998, p.\ 201

\bibitem[1966]{Warner66} Warner B., 1966, Nat 211, 55

\bibitem[1974]{Wegner74}  Wegner G., Petford A.D., 1974,
                MNRAS 168, 575

\bibitem[1976]{Wolff76} Wolff S.C., Hagen W., 1976,
                PASP 88, 119

\end{thebibliography}
\end{document}